\documentclass[aps,prl,twocolumn,groupedaddress,showpacs,longbibliography]{revtex4-1}

\usepackage{amssymb,amsmath}
\usepackage{graphicx}
\usepackage[english]{babel}

\begin{document}


\title{Revealing the topology of quasicrystals with a diffraction experiment}

\author{A. Dareau$^{1,\dagger}$}
\author{E. Levy$^{2,3,\dagger}$}
\author{M. Bosch Aguilera$^{1}$}
\author{R. Bouganne$^{1}$}
\author{E. Akkermans$^{2}$}
\author{F. Gerbier$^{1}$}
\author{J. Beugnon$^{1}$}

\email[]{beugnon@lkb.ens.fr}

\affiliation{$^1$Laboratoire Kastler Brossel,  Coll\`ege de France, CNRS, ENS-PSL Research University, UPMC-Sorbonne Universit\'es, 11 Place Marcelin Berthelot, 75005 Paris, France}
\affiliation{$^2$Department of Physics, Technion Israel Institute of Technology, Haifa 32000, Israel}
\affiliation{$^3$Rafael Ltd., P.O. Box 2250, Haifa 32100, Israel}

\date{\today}
\begin{abstract}
Topological properties of crystals and quasicrystals is a subject of recent and growing interest. This Letter reports an experiment where, for certain quasicrystals, these properties can be directly retrieved from diffraction. We directly observe, using an interferometric approach, all the topological invariants of finite-length Fibonacci chains in their diffraction pattern. We also demonstrate quantitatively the stability of these topological invariants with respect to structural disorder.
\end{abstract}

\maketitle

Concepts issued from topology, a well-established branch of mathematics, find increasing use in various areas of physics. Topology generalizes the notion of symmetry classes by classifying physical objects into distinct families, or topological classes, which cannot be related by continuous deformations. A well-known example is the Gauss classification of surfaces in three-dimensional space by their genus, roughly equivalent to the number of holes piercing them ({\it e.g.} a doughnut cannot be continuously deformed into a sphere).  Topological classes are characterized by a set of integer numbers called topological invariants due to their stability against a broad range of perturbations. A celebrated example of such invariant is the Chern number which characterizes the quantization of conductance for the integer quantum Hall effect \cite{Klitzing86}. Topological invariants play an important role in many other situations, including the study of topological defects in symmetry-broken phases \cite{Mermin79}, anomalies in quantum field theory \cite{nielsen83}, topological insulators and superconductors \cite{Hasan10}, band structures with Dirac \cite{CastroNeto09} or Weyl points \cite{Xu15,Lu15} and also quasicrystals. 

\begin{figure}[hbt!]
\centering
\includegraphics[width=0.45\textwidth]{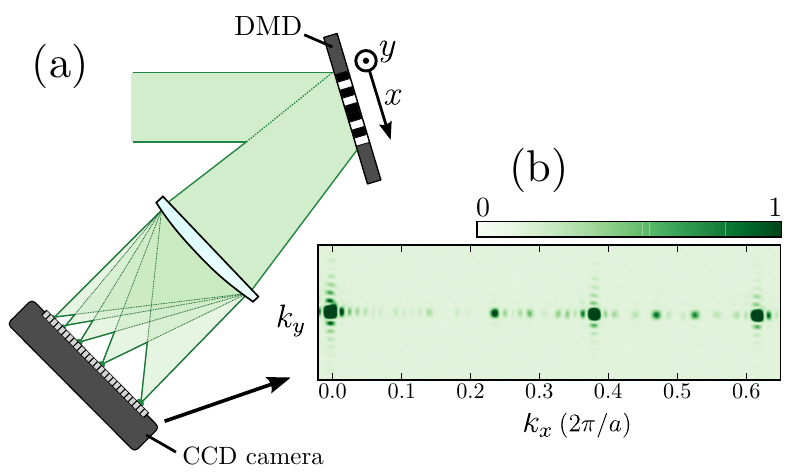}
\caption{Experimental setup. (a) Principle of the experiment. A collimated laser beam is sent to a diffraction grating programmed on a digital micromirror device (DMD). We measure the far-field diffraction pattern on a CCD camera \cite{REFSM}. (b) Characteristic diffraction pattern for a grating following a Fibonacci sequence along the horizontal $x$ direction (and uniform along the vertical $y$ direction).}\label{fig1}
\end{figure}

Quasicrystals (hereafter QC) are structures having some of their physical properties modulated according to a deterministic but non-periodic pattern. Despite their lack of periodicity, QC exhibit long-range order and show sharp diffraction peaks \cite{Shechtman84,Steinhardt87,Katz86,Yamamoto96}. Quasi-periodic systems have been well investigated, especially in one dimension \cite{Simon82,Kunz86,Damanik11}. The study of spectral characteristics of propagating waves (acoustic, optical, matter, ...) in these quasi-periodic structures reveals a highly lacunar fractal energy spectrum, with an infinite set of energy gaps \cite{Kohmoto87,Luck89,Tanese14}. The gap labelling theorem provides a framework for the topological classification of these gaps and plays for quasi-periodic media a similar role to that of Bloch theorem for periodic ones \cite{bellissard89,Bellissard92}. Bloch theorem allows to label the eigenstates of a periodic system with a quasi-momentum and to identify topological (Chern) numbers expressed in terms of a Berry curvature. This labelling is robust as long as the lattice translational symmetry is preserved. Similarly, the gap labelling theorem allows to associate integer-valued topological numbers to each gap in the spectrum \footnote{These topological numbers are K-theory invariants. They are not strictly speaking Chern numbers which are invariants describing the topology of smooth Riemannian manifolds. The Fibonacci chain, and quasicrystals in general, cannot be ascribed such a smooth manifold. However there may exist an interpolation between both situations that could establish a link between Chern numbers and the topological numbers we measure in this work \cite{Kraus12,Gitelman17}.}. Those integers can be given both a topological meaning and invariance properties akin in nature to Chern numbers but not expressible in terms of a curvature \cite{bellissard89,Belissard93,Bellissard92}. In both cases the topological invariants are related to the energy spectrum and remain the same when the Hamiltonian is perturbed, as long as the gaps do not close.
\begin{figure*}[htb!!]
\centering
\includegraphics[width=0.8\textwidth]{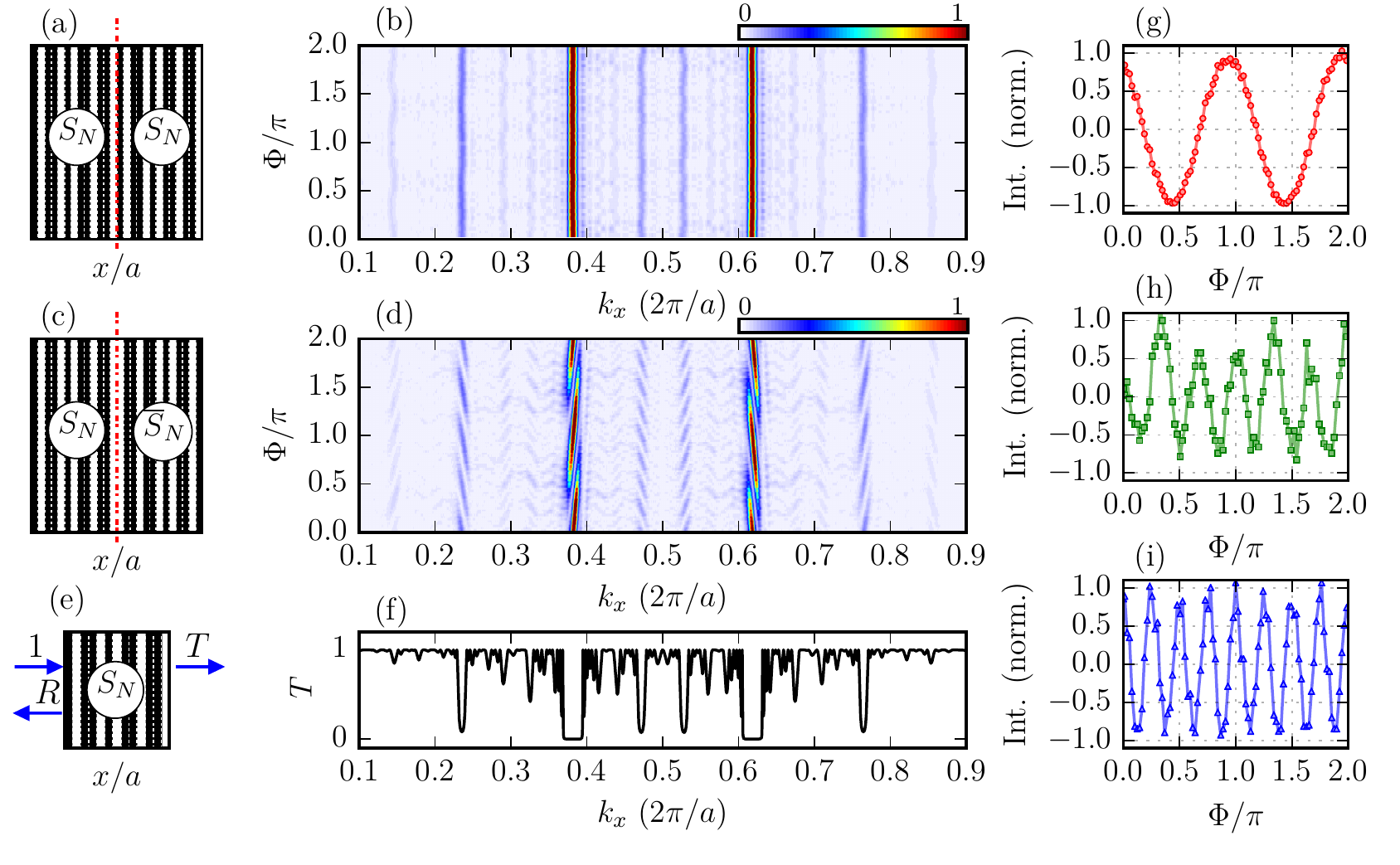}
\caption{Influence of the phason degree of freedom on the diffraction pattern of a Fibonacci grating.
(a) Structure composed by the juxtaposition of two identical Fibonacci chains along $x$ (repeated along $y$). White lines correspond to reflecting mirrors (B letters in the Fibonacci chain) and black to non-reflecting mirrors (A letters). We show for clarity a chain of length $F_6=13$. 
(b) Measured diffraction pattern for two side-to-side Fibonacci chains  of length $F_{10}=89$. Each line corresponds to the diffraction pattern for a given value of the phason degree of freedom $\Phi$. 
(c) Structure composed by the juxtaposition of a Fibonacci chain and of the reversed chain along $x$ (repeated along $y$).
(d) Measured diffraction pattern for the structure depicted in Fig.\,2(c) with length \textbf{$2\times F_{10}=89$}. 
(e) Sketch of a transmission experiment.
(f) Transmission of a 1D Fibonacci chain as calculated in \cite{Levy15}. Gaps appear at the same position as the peaks of the diffraction pattern.
(g)-(i) Cuts along vertical lines of Fig.\,2(d) for three values of $k_x \approx k_q\approx(2\pi/a) \times (0.618 $ (g), 0.146 (h) and 0.472 (i)) corresponding respectively to topological numbers $q=1,-3$ and $4$.}\label{fig2}
\end{figure*}

The purpose of this Letter is to present an experiment in which we show that topological numbers can be obtained from a purely structural aspect of QC and directly measured in their diffraction pattern. We observe that the diffraction amplitude for each peak of the diffraction pattern displays a well-defined winding number, which coincides with the integer gap label in the spectrum at the corresponding wavevector. We also reveal experimentally the robustness of these topological invariants against structural noise in the chain. Our method provides a novel experimental approach to study topological properties of matter, different from transport or topological pumping experiments  \cite{Klitzing86,Hasan10,Kraus12b}, and from spectral measurements \cite{Tanese14,Baboux17,Skirlo15}.

In this work, we focus on the iconic example of QC in 1D, the Fibonacci chain. This refers to a one-dimensional structure modulated according to the Fibonacci sequence \cite{Kohmoto87}. The gap-labelling theorem \cite{bellissard89,Bellissard92,Belissard93} applied to the infinite Fibonacci chain (denoted as $S_\infty$) states that the integrated density of states $\mathcal{N}$ for a wave number $k$ inside a gap takes the simple form
\begin{equation}
\mathcal{N}\left(k_{\rm in\,\, gap}\right)=p+q/\tau.\label{eq1}
\end{equation}
where $p,q \in \mathbb{Z}^2$, $p(q)$ keeps $\mathcal{N}$ normalized within $[0,1]$ and $\tau=(1+\sqrt{5})/2$ is the golden mean. The integers $p,q$ are topological invariants, \textit{i.e} the relation (\ref{eq1}) holds irrespective of the precise form of the Hamiltonian as long as the gaps remain open. These features have been recently revisited \cite{Dana14,Tanese14,Levy15} in the wake of the growing interest in topological properties of solid-state systems. 

In our experiment (illustrated in Fig.\,1), we realized Fibonacci chains using a Digital Micromirror Device (DMD), {\it i.e.} an array of about one million micron-sized mirrors (``pixels'') of size $a \times a$. Each mirror can be independently switched between a reflective ($B$) and a non-reflective ($A$) state. We illuminated the generated grating with monochromatic light \cite{REFSM}, and observed the far-field diffraction pattern on a CCD camera. According to Fourier optics, the far-field diffraction amplitude is determined by the Fourier transform of the reflectance of the DMD,  $ \mathcal{A}(k_x)=\sum_{\{ x_B \}} e^{i k_x x_B}$, where $\{ x_B \}$ denotes the configuration of reflecting pixels (multiplied by an envelope function originating from the DMD structure). The measured light intensity $I$ is proportional to $\vert \mathcal{A}(k_x) \vert^2$.

\begin{figure*}[htb!!!]
\centering
\includegraphics[width=0.9\textwidth]{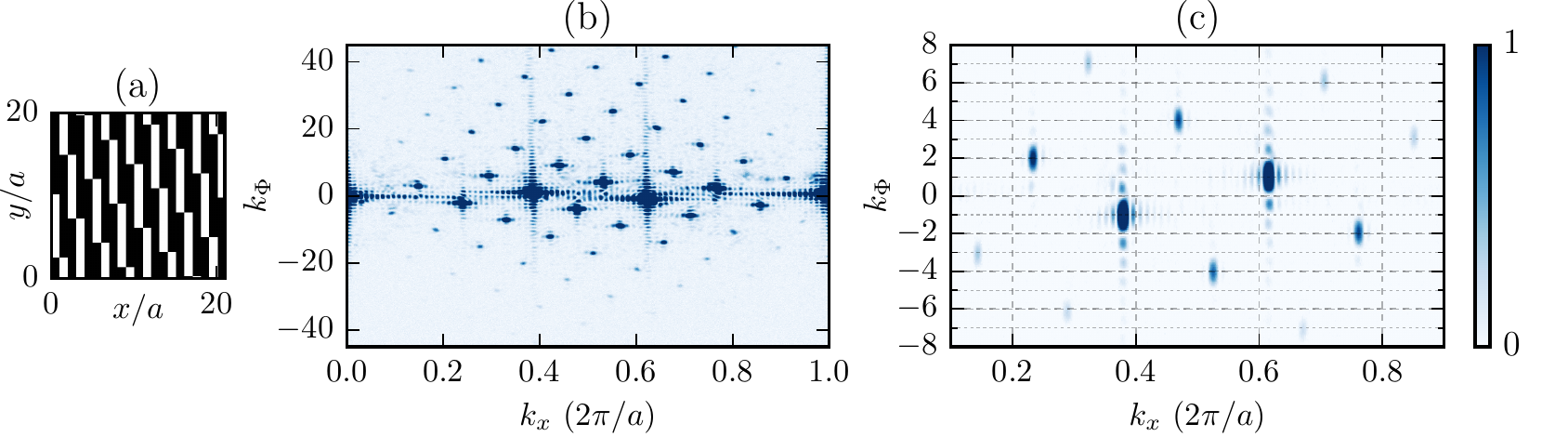}
\caption{Two-dimensional map of topological numbers. (a) Structure of the diffraction grating. White (resp. black) pixels correspond to reflective (resp. non-reflective) mirrors. The bottom line is the Fibonacci sequence for $\Phi=0$. The other lines correspond to the phase-shifted Fibonacci chains scanning $\Phi$ from $0$ to $2\pi$. The length of each chain in this illustration is $F_7=21$ for clarity. (b) Two-dimensional diffraction pattern for 89 phason-shifted chains with $F_{10}=89$ letters.  We observe peaks at the same positions along $k_x \approx k_q$ as in Fig.\,2(b) but shifted along the vertical axis to $k_\Phi =q$. (c) Zoom on Fig.\,3(b). }\label{fig3}
\end{figure*}

We programmed the pixel configurations \textbf{$\{ x_A,x_B \}$} to follow one of the possible Fibonacci sequences with fixed, finite length. Finite-size chains of given length form a family of one-dimensional quasi-periodic structures, which are all segments of the infinite chain $S_\infty$, and which can be deduced from each other by changing a structural degree of freedom  termed a \textit{phason} \cite{Levy15}. To generate the chains $S_N$, we consider a discrete map $\chi_n$ taking, for each pixel $n$, two possible values $\pm 1$ that we associate to the state of the mirror $\left(A =1,B=-1\right)$.  We use, among possible choices \cite{Kesten66,Ostlund84,Godreche90}, the characteristic function \cite{Kraus12},
\begin{equation} 
\chi_n (\Phi) = {\rm sign} \, \left[ \cos \left(\frac{ 2 \pi n }{ \tau} + \Phi + \Phi_0 \right) - \, \cos \left( \frac{ \pi }{ \tau} \right) \right],
\label{chi}
\end{equation}
where the phason $\Phi$ appears as a continuous and $2\pi$-periodic variable. Each finite segment is defined by $S_N (\Phi)  \!\!\equiv\!\! \left[ \chi_1 \, \!\chi_2 \,\! \cdots \! \chi_{F_N} \right]$ (here $F_N$ denotes the $N^{\rm th}$ Fibonacci number). For a given structure of length $F_N$, varying $\Phi$ over a period $[0,2\pi)$ induces a series of $F_N$ local structural changes occurring one at a time and generating $F_N$ non-redundant chains (see Fig.\,3). The value of $\Phi_0$ in (\ref{chi}) is chosen as $\Phi_0=- (F_N+1)\pi / \tau$ so that the chain $S_N(0)$ presents a mirror symmetry with respect to its center.

As a preliminary experiment, we studied a Fibonacci grating of length $L=F_{10}\times a$, with $F_{10}=89$. The pattern, shown in Fig.\,1, consists of one-pixel-large vertical lines either in reflective ($B$) or non-reflective ($A$) state according to Eq.\,\eqref{chi}. We show in Fig.\,1(b) the measured diffraction pattern. We observe many diffraction peaks at wavenumbers $k_x \approx k_q$, where $k_q= p+q/\tau$ (in units of $2\pi/a$) denote the expected positions for the infinite chain $S_\infty$ \cite{Steinhardt87} \footnote{Although $p,q$ are the same integer numbers as the gap labels for an infinite Fibonacci chain, the former are not topologically protected and the peak position can change if the chain deviates from the ideal case, unlike the density of states in Eq.\,\ref{eq1} (see  \cite{REFSM})}.  In the following, we note $\mathcal{A}_0(k_x)$ the diffraction amplitude from this reference chain and $I_0(k_x)$ the corresponding intensity. 

To reveal the topological features hidden in this pattern, we consider the $F_N$ distinct chains obtained with different values of $\Phi$ for a given $N$. In a first experiment, we measured the diffraction pattern of a grating consisting of a Fibonacci chain next to itself (``$S_N+S_N$'' configuration, see Fig.\,2(a)). All the results are consolidated to form the graph shown in Fig.\,2(b), where each $\Phi$ value corresponds to a single diffraction measurement. We observe vertical stripes located at a value $k_x \approx k_q$ corresponding to the position of spectral gaps (Fig.\,2(f)). These stripes are independent of $\Phi$. In a second experiment, we used a grating consisting of a Fibonacci chain next to its mirror image (``$S_N+\overline{S}_N$'' configuration, see Fig.\,2(c)). The resulting diffraction pattern shown in Fig.\,2(d) is strikingly different. The vertical stripes are now striated to form a regular and well-structured pattern. The intensities, measured at the original location of the diffraction peaks $k_q$, vary sinusoidally with $\Phi$ (Fig.\,2(g),(h) and (i)), with a period that we identify as $\pi/|q|$.  The winding number $q$ is identical to the topological gap label associated with each peak in Eq.\,(\ref{eq1}): the ``velocity'' at which the peak is crossed is $\vert q \vert$ and the crossing direction is determined by the sign of $q$. 

We explain these observations in terms of a ``Young's double slit'' interference between the waves diffracted by the two chains. In both experiments, the diffracted field is the coherent sum of the amplitude for each chain. Changing $\Phi$ is equivalent, as discussed in \cite{REFSM}, to a  translation along the infinite chain $S_\infty$ and thus to an additional phase factor in the diffracted amplitude. We show in the Supplemental Material \cite{REFSM} that, for a given wavenumber $k_q$, the phase shift of the diffracted field with respect to the reference chain  is $q  \Phi$. In the first experiment (``$S_N+S_N$'' configuration -- Fig.\,2(a)), the diffraction amplitudes from each chain at $k_x \approx k_{q}$ are respectively proportional to $e^{i(q\Phi+\phi_s)}$ and to $e^{i q \Phi}$, where the phase $\phi_s=k_q a F_N$ is due to the separation between the two ``slits". The intensity $I_{S_N+ S_N} = 4 I_0 (k_q) \cos^2(\phi_s/2)$ is thus independent of $\Phi$. By contrast, in the second experiment (``$S_N+\overline{S}_N$'' configuration -- Fig.\,2(c)), the two chains $S_N$ and $\overline{S}_N$ are related by a  mirror symmetry, and their respective diffraction amplitudes are proportional to  $e^{i  (q\Phi+\phi_s)}$ and $e^{-i(q\Phi+\phi_s)}$ \cite{REFSM}. This leads to 
\begin{equation} 
I_{S_N+\overline{S}_N} (k_{q},\Phi) =4 I_0(k_{q})\cos^2\left( q \Phi+\phi_s\right) \, ,
\label{IKpq}
\end{equation}
a sinusoidal function of $\Phi$ with period $\pi/|q|$\textbf{,} as observed experimentally \footnote{Note that the topological information is thus contained within only half the $2\pi$-period of $\Phi$.}.

In a third experiment, we used the $e^{i q \Phi }$ dependence of the diffraction amplitude to obtain a single-shot measurement of all possible topological numbers $q$. We programmed each of the $F_N$ different $S_N (\Phi)$ chains on horizontal lines along the $x$-direction, thereby forming an $F_N\times F_N$ array where the vertical direction $y$ can be identified with the phason $\Phi$, taking values from $0$ to $2\pi$ when $y$ goes from 0 to $aF_N$. The measured diffraction pattern in the $(k_x,k_y)$ plane is associated to a pattern in the $(k_x,k_\Phi)$ plane with a normalization of the $k_y$ axis by \textbf{$2\pi/(a F_N)$}. The diffraction pattern of the $F_N \times F_N$ array,  shown in Fig.\,3(b,c), displays a set of peaks at well-defined positions in the $(k_x,k_\Phi)$ plane. The $k_x$ coordinate is the same as observed previously for single chains, while the $k_\Phi$ coordinate is equal to the topological number $q$. This result follows from our previous analysis,  \textsl{i.e.} from the $e^{i q \Phi}$ dependence of the diffracted field. This direct measurement of topological numbers thereby provides a ``topological map'' of the Fibonacci chain in a single-shot experiment. We observe an asymmetry between the intensity of the diffraction peaks for positive and negative $k_\Phi$ values. This feature is due to the internal structure of the DMD, which leads to  an asymmetric diffraction envelope, and is not an intrinsic property of Fibonacci chains.

In the Supplemental Material \cite{REFSM}, we propose a construction based on the ``Cut and Project'' method to interpret further the experimental results of Fig.\,3. We show that the reciprocal space of the ($x,\Phi$) phase space has the topology of a torus. The topological numbers associated to each diffraction peak are winding numbers around the torus. We find the exact positions of the diffraction peaks, which depend on the chain length, and show that the corresponding topological numbers are exact, in the sense that they take the \textit{same} integer value irrespective of the chain length. The finite length $F_N$ only results in the existence of an upper limit to the topological numbers, $0< \vert q \vert \leq F_N/2$. Experimentally, we observe all the possible values for $F_{10}=89$.
\begin{figure}[htb!!!!]
\centering
\includegraphics[width=0.45\textwidth]{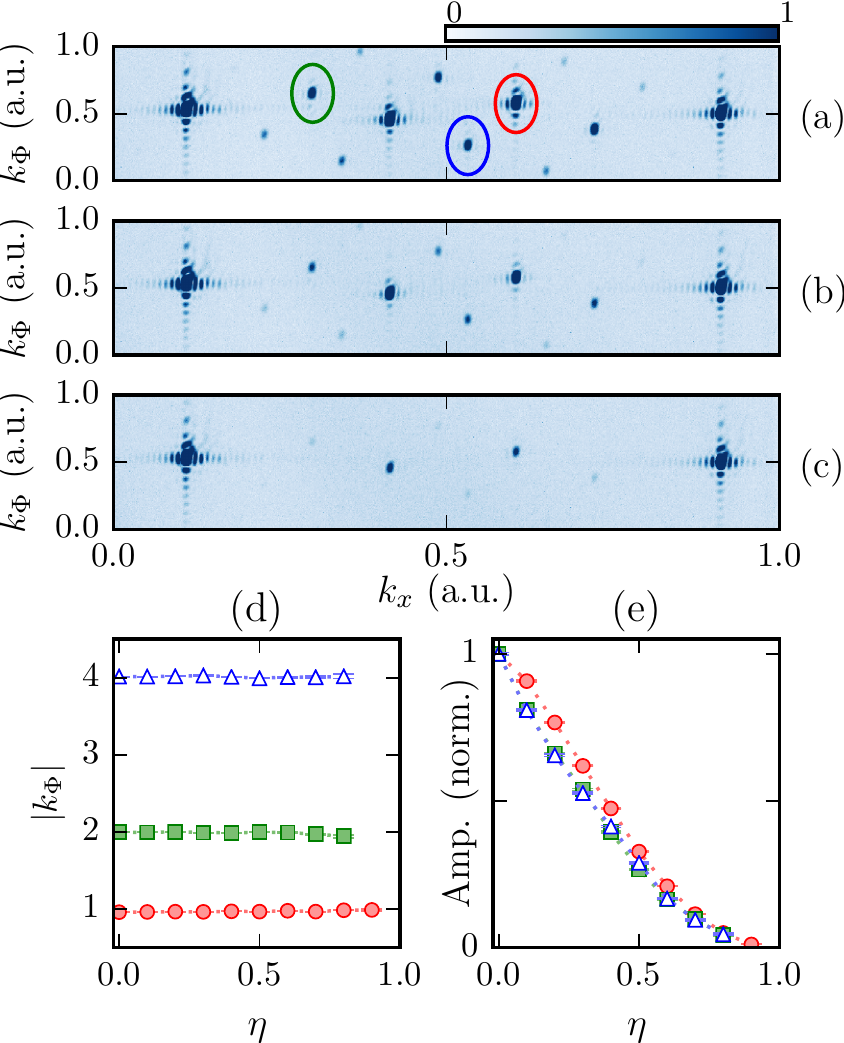}
\caption{Robustness of the topological features against structural noise.  (a-c) Examples of 2D diffraction patterns for noise levels $\eta=0.1, 0.5$ and $0.8$. (d) Position of the diffraction peaks along the $k_\Phi$ axis for  $q=1$ (circles), $q=2$ (squares) and $q=-4$ (triangles). (e) Corresponding amplitude of the diffraction peaks normalized to the amplitude of the $q=1$ peak without noise. Statistical error bars in (d-e) are about the size of the markers.}
\label{fig4}
\end{figure}

Finally, in a fourth experiment, we explored the robustness of the diffraction pattern and of its topological features against structural noise. We studied  a similar configuration as in the third experiment described above, but introduced noise by randomly selecting a fraction $\eta$ of the mirrors ($0 \leq \eta \leq 1$) whose states (reflective or non-reflective) were also chosen at random. Thus, $\eta=0$ corresponds to a non-perturbed pattern and $\eta=1$ to a fully random pattern. The resulting diffraction signal is averaged over many realizations of the noise \cite{REFSM}. In Fig.\,4(a-c), we show how the two-dimensional diffraction pattern evolves with increasing $\eta$. As expected, peaks are progressively washed out when increasing the fraction $\eta$. We select in Fig.\,4(d-e) three specific peaks ($q=1$, $q=2$ and $q=-4$) and show the relative evolution of the peak positions and normalized amplitude when varying $\eta$. The peak position is almost not influenced by the presence of noise whereas the peak amplitude decreases strongly when increasing the noise level. Further studies of the influence of noise on the experimental configurations of Fig.\,2 are reported in \cite{REFSM}. These results demonstrate the robustness of the topological properties of the Fibonacci chains captured by the winding numbers of the diffraction amplitudes.

To conclude, we have presented a novel but simple approach to measure topological invariants associated with quasicrystalline structures. Although we worked with the Fibonacci chain, our method is not restricted to it and can be applied to a broader class of quasicrystals. A two-dimensional array of mirrors is also well suited to extend this study to the determination of topological invariants for two-dimensional tilings where much less is known than for one-dimensional structures \cite{Kraus13,Tran15}.

\begin{acknowledgments}
$^\dagger$A.D. and E.L. contributed equally to this work. We  acknowledge  financial  support from the European Research Council under grant 258521 (MANYBO) and by the Israel Science Foundation under grant 924/09. We thank R. Mosseri for stimulating discussions.
\end{acknowledgments}

\bibliography{fibobib}

\begin{thebibliography}{44}
\bibitem [{REF()}]{REFSM}%
  \BibitemOpen
  \href@noop {} {\enquote {\bibinfo {title} {See {S}upplemental {M}aterials for
  a description of experimental methods and complementary calculations, which
  includes refs.
  \cite{Levy16,Konig97,Konig00,Parameswaran13,bruijn81,elser86,zia85,mosseri92}},}\
  }\BibitemShut {NoStop}%
\end{thebibliography}

\end{document}